\documentclass[english,prl,twocolumn,amsmath,amssymb,superscriptaddress,floatfix]{revtex4}
\usepackage[T1]{fontenc}
\usepackage[latin1]{inputenc}
\usepackage{graphicx}

\usepackage{babel}
\makeatother

\begin{document}

\title{Atomic force microscope nanolithography of graphene: cuts, pseudo-cuts and tip current measurements.}

\author{R.K. Puddy}
\affiliation{Cavendish Laboratory, University of Cambridge, Cambridge CB3 0HE, United Kingdom}

\author{P.H. Scard}
\affiliation{Cavendish Laboratory, University of Cambridge, Cambridge CB3 0HE, United Kingdom}

\author{D. Tyndall}
\affiliation{Cavendish Laboratory, University of Cambridge, Cambridge CB3 0HE, United Kingdom}

\author{M.R. Connolly}
\affiliation{Cavendish Laboratory, University of Cambridge,
Cambridge CB3 0HE, United Kingdom}

\author{C.G. Smith}
\affiliation{Cavendish Laboratory, University of Cambridge, Cambridge CB3 0HE, United Kingdom}

\author{G.A.C. Jones}
\affiliation{Cavendish Laboratory, University of Cambridge, Cambridge CB3 0HE, United Kingdom}

\author{A. Lombardo}
\affiliation{Engineering Department, University of Cambridge, Cambridge CB3 0FA, United Kingdom}

\author{A.C. Ferrari}
\affiliation{Engineering Department, University of Cambridge, Cambridge CB3 0FA, United Kingdom}

\author{M.R. Buitelaar}
\altaffiliation{mrb51@cam.ac.uk}
\affiliation{Cavendish Laboratory, University of Cambridge, Cambridge CB3 0HE, United Kingdom}

\begin{abstract}

We investigate atomic force microscope nanolithography of single and
bilayer graphene. In situ tip current measurements show that cutting
of graphene is not current driven. Using a combination of transport
measurements and scanning electron microscopy we show that, while
indentations accompanied by tip current appear in the graphene
lattice for a range of tip voltages, real cuts are characterized by
a strong reduction of the tip current above a threshold voltage. The
reliability and flexibility of the technique is demonstrated by the
fabrication, measurement, modification and re-measurement of
graphene nanodevices with resolution down to 15 nm.

\end{abstract}

\maketitle

Scanning probe microscopy, as well as being a powerful tool for
imaging and spectroscopy, has also shown great potential for the
manipulation and patterning of materials on the nanometer scale
\cite{Tseng2005, Tapaszto}. Atomic force microscope (AFM)
nanolithography, in particular, is now routinely used for the
fabrication of quantum dots and quantum wires in materials such as
Si and GaAs \cite{Campbell1995, Keyser2000}. AFM nanolithography
also has significant potential for device fabrication in graphene, a
material of intense current interest due to its exceptional
mechanical, electronic and optical properties \cite{Geim2007,
Bonaccorso2010}. Most commonly, graphene nanoscale devices are
fabricated using conventional electron-beam lithography and
subsequent plasma etching
\cite{Han2007,Ponomarenko2008,Stampfer2008}. AFM lithography offers
several advantages over electron-beam lithography: it has higher
ultimate resolution, can be performed under ambient conditions and
allows in situ device measurement and modification.

Usually AFM nanolithography is performed in air at room temperature.
Under these conditions a water meniscus forms between the AFM tip
and the substrate. The presence of an electric field, resulting from
the voltage between the tip and substrate, dissociates water into
hydrogen (H$^+$) and hydroxyl (OH$^-$) ions. When the voltage on the
tip is negative with respect to the substrate the hydroxyl ions
oxidize the graphene surface, creating the desired nanostructure.
Several important factors determine the reliability and resolution
of AFM lithography such as the applied tip voltage (or electric
field strength), the humidity, tip velocity, applied force, and the
conductivity of the substrate \cite{Tseng2005}. This process is now
well understood for a variety of semiconductors and metals. However,
in the case of graphene many of the key parameters have not been
well established and device fabrication by AFM lithography is not
yet routine. For example, the necessary threshold tip voltage for
graphene oxidation reported in the literature varies in magnitude
between $\sim$ -5 V \cite{Neubeck2010} and -35 V
\cite{Weng2008,Masubuchi2009} and in one report oxidation could only
be initiated from a graphene edge \cite{Giesbers2008}. Moreover,
there has been no systematic study of the tip current during AFM
lithography of graphene.

Here, we investigate in detail the cutting of the graphene lattice
with an AFM tip. In particular, we measure the tip current,
I$_{tip}$, during the cutting process and find that we cut graphene
only when I$_{tip}$ drops below our noise floor. We also find that
pseudo-cuts appear when I$_{tip}$ is non-zero. These pseudo-cuts, in
which the electron system of graphene remains intact, cannot
reliably be distinguished from real cuts by AFM height imaging.
However, the differences between real and pseudo-cuts become
apparent using transport experiments and scanning electron
microscopy (SEM). This ability to distinguish between real and
pseudo-cuts is crucial for device fabrication in graphene.

\begin{figure*}
\includegraphics[width=175mm]{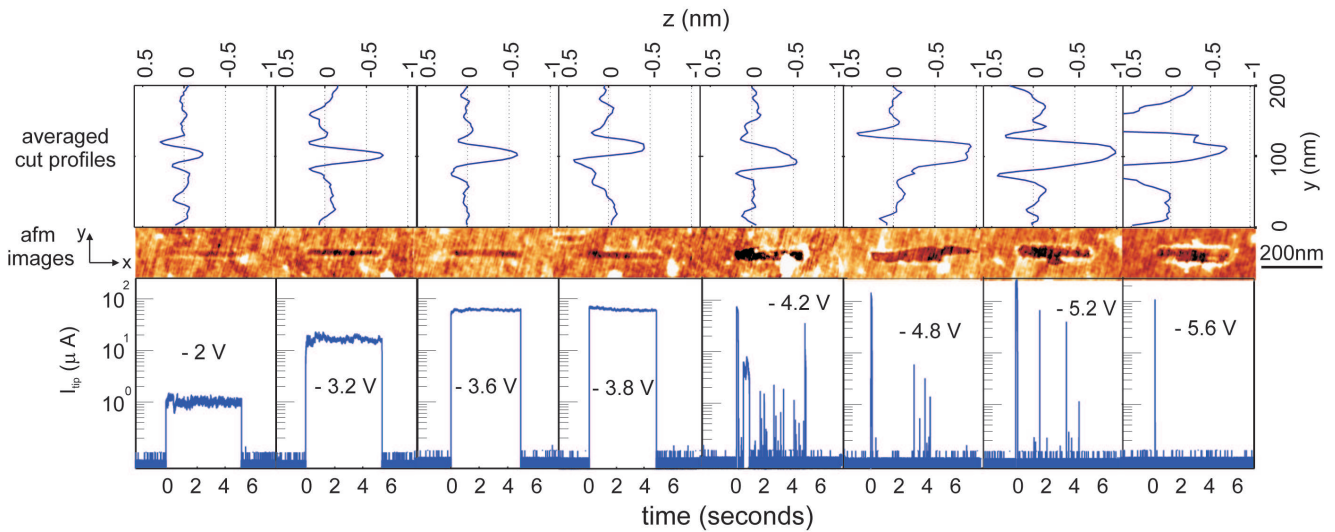}
\caption{\label{Fig1} (color online) A series of cuts or
indentations in a single layer graphene flake, made using an AFM tip
in contact mode at increasingly negative voltages relative to the
grounded flake. The upper panels show the averaged height profile
cross-sections of the cuts, the central panels show the
corresponding AFM micrographs and the bottom panels show the
current, I$_{tip}$, through the AFM tip as a function of time $t$,
recorded during the cutting process where $t = 0$ is the start of
tip contact. The series is performed using a single, non coated,
doped silicon tip \cite{tips}. Tip speed was 50 nm/s and the
relative humidity was $\sim 50\%$.}
\end{figure*}

To investigate the voltage and current dependence of AFM
nanolithography on graphene we use a Veeco Dimensions 3100 AFM
system with non-coated, doped silicon tips for both imaging and
lithography \cite{tips}. Imaging is performed in tapping mode and
lithography carried out in contact mode. Single (SLG) and bilayer
(BLG) graphene flakes are produced by micro-mechanical exfoliation
on 300 nm SiO$_{2}$ with a highly doped Si back gate. Optical
\cite{Casiraghi2007} and Raman spectroscopy \cite{Ferrari2006} are
used to assess the layer number and quality. All flakes are
electrically contacted and characterized at room temperature.
Contacts are defined in polymethylmethacrylate (PMMA) resist by
electron beam lithography and metalized with Ti/Au (5/50 nm) by
evaporation and lift off.

Samples are annealed for $\sim$ 10 mins. at 300$^{o}$C in forming
gas prior to AFM lithography, which we find to be a crucial step for
reliable cutting of our flakes as it removes contamination which can
prevent oxidation. During AFM nanolithography, we use a 50 nm/s
scanning speed and a relative humidity of around 50$\%$. The
lithographically defined trenches vary in width from 15 nm up to 100
nm with typical values of 30 nm. We find that the widths depend only
weakly on scanning speed and humidity, with cuts slightly wider with
decreasing scanning speed and increasing relative humidity.
Individual tip characteristics, most likely tip radius, appear to be
more important.

We firstly investigate the current through the biased AFM tip during
lithography. Figure \ref{Fig1} shows a series of cuts performed on a
SLG at various tip voltages. The AFM tip is negatively biased with
respect to the flake, which is grounded via a Stanford SR570 current
preamplifier. The lower panels of Fig.~\ref{Fig1} show the current,
I$_{tip}$, through the tip to ground, as a function of time where $t
= 0$ is the time at which the tip bias V$_{tip}$ is applied and the
tip starts along its predefined path. Above this is the
corresponding AFM image of the scanned area. The upper cells show
the averaged height cross section across the cuts. Indentations
begin to appear on the graphene surface, accompanied by finite
$I_{tip}$, at around V$_{tip}$ = -2 V. The tip current then drops to
$\sim$ 0 above a threshold, V$_{thresh}$. Threshold voltages for our
tips vary from $\sim$ -3.5 V to $\sim$ -5 V. Trenches created at the
smallest voltages occasionally disappear over the course of hours or
days, regaining their original shape. Trenches created at larger
voltages remain unchanged after several weeks. We also note that, in
both regimes, ridges are frequently formed along the trench edges
where the electric field is lower. This may be due to the formation
of stable oxides similar to that reported in Refs.
\cite{Weng2008,Neubeck2010}.

\begin{figure}
\includegraphics[width=85mm]{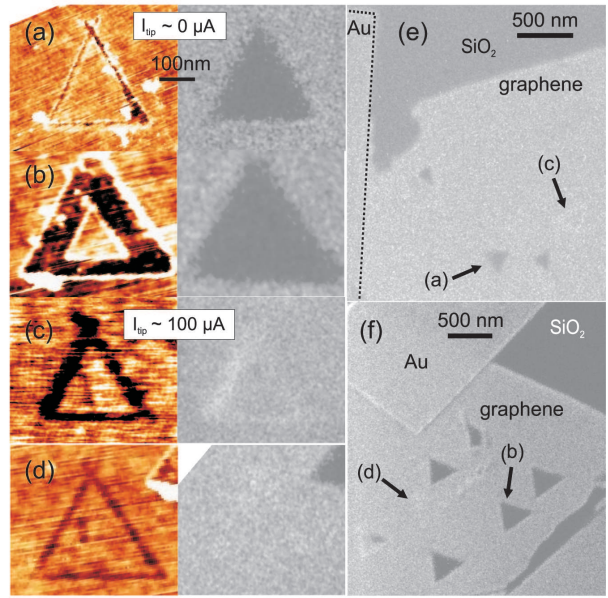}
\caption{\label{Fig2} (color online) (a,b) AFM (left) and SEM images
(right) of two triangles cut with |V$_{tip}$| > |V$_{thresh}$| such that the tip current,
I$_{tip} \sim 0$ during cutting. The central regions are clearly visible in the topographic
AFM images but absent in the SEM. (c,d) AFM (left) and SEM images (right) of two triangles
cut with |V$_{tip}$| < |V$_{thresh}$|
such that the current through the tip during cutting is $\sim$ 100 $\mu$A.
The AFM images are qualitatively similar to those in panels (a) and (b) but no contrast
is seen in the SEM images. (e,f) SEM images of the areas of
the graphene flake on which the triangles were cut. The arrows indicate the
locations of the triangles shown in panels (a)-(d).}
\end{figure}

In order to investigate the nature of the marks created in the two
regimes we cut triangles into a SLG with V$_{tip}$ both above and
below V$_{thresh}$ and imaged them using both AFM and SEM. Figures
\ref{Fig2}(a) and \ref{Fig2}(b) show two triangles, cut with
|V$_{tip}$| > |V$_{thresh}$| and thus I$_{tip} \sim$ 0, imaged using
AFM (left image) and SEM (right image). The central regions of the
triangles, clearly visible in the AFM image, are significantly
darker in the SEM images as compared to the bulk. Figures
\ref{Fig2}(c) and \ref{Fig2}(d) show two triangles, cut with
|V$_{tip}$| < |V$_{thresh}$| and I$_{tip} \sim$ 100 $\mu$A, again
imaged using AFM (left) and SEM (right). Though the AFM images are
qualitatively similar to those in Figs.~\ref{Fig2}(a) and
\ref{Fig2}(b), the triangles are barely visible in the SEM images
with the central regions showing no contrast with the bulk. In all
cases, SEM imaging is carried out using an accelerating voltage of
500 V. Note that for these low acceleration voltages (i.e. below
$\sim 1$ kV), the contacted graphene is easily visible on the
SiO$_2$ substrate. This is illustrated in Figs.~\ref{Fig2}(e) and
\ref{Fig2}(f) which show SEM images of the areas of the flake on
which the triangles were cut. The arrows indicate the locations of
the triangles in Figs.~\ref{Fig2}(a)-\ref{Fig2}(d). The strong
contrast of the graphene flakes on the SiO$_2$ substrate is
attributed to differences in the surface electrostatic potential
between the bare SiO$_2$ substrate and the regions covered by the
(electrically contacted) graphene, similar to that observed for
carbon nanotubes \cite{Brintlinger}. This immediately allows us to
conclude that the triangles shown in Figs.~\ref{Fig2}(a) and
\ref{Fig2}(b), which appear dark in the SEM images, are electrically
isolating, while those of Figs.~\ref{Fig2}(c) and \ref{Fig2}(d) are
electrically connected to the bulk. This behaviour is consistent
over all 10 pairs of triangles measured.

For further confirmation, the tip is placed inside the triangles and
voltages below threshold are applied. We find that we never measure
current above our noise floor with triangles cut with |V$_{tip}$| >
|V$_{thresh}$| while we always measure current with triangles cut
with |V$_{tip}$| < |V$_{thresh}$|. Furthermore, the current measured
from within these triangles is no smaller than the current measured
from outside the triangles. The resulting indentations can be seen
as the short lines within Figs.~\ref{Fig2}(c) and ~\ref{Fig2}(d)
whereas no such marks are seen in Figs.~\ref{Fig2}(a) and
\ref{Fig2}(b). We conclude that electrically isolating cuts are made
only for negative tip voltages larger than |V$_{thresh}$| at which
point the applied electric field is sufficiently strong to initiate
the oxidation process. We believe that for tip voltages below the
threshold, the SLG is merely pressed into contact with the SiO$_{2}$
surface by displacing water trapped between graphene and the
substrate. The graphene in this case remains unbroken and
electrically conducting.

Our findings may help to clarify previous measurements on highly
oriented pyrolytic graphite (HOPG) in which an AFM was used for
local oxidation \cite{Kim2003,Park2007,Jiang2008}, the
interpretations of which are conflicting. While some reports
identify a strong Fowler-Nordheim field emission dependence of the
tip current on the applied voltage, attributing the oxidation to a
field emission enhanced chemical reaction \cite{Kim2003}, other work
on HOPG finds that holes are only observed in the absence of a tip
current \cite{Park2007,Jiang2008}. Our observations are consistent
with the latter and indicate that the presence of a large current
points to a failure of graphene oxidation. This excludes cutting
mechanisms such as heating of the carbon lattice. The strong
reduction of the tip current during cutting can be understood if
graphene oxidation occurs once the local electric field strength
exceeds the reaction activation energy. As graphene is oxidized
below the tip, the tip-graphene contact is disrupted and only the
very small electron flow of the reaction electrons remains
\cite{resolution}.

\begin{figure}
\includegraphics[width=85mm]{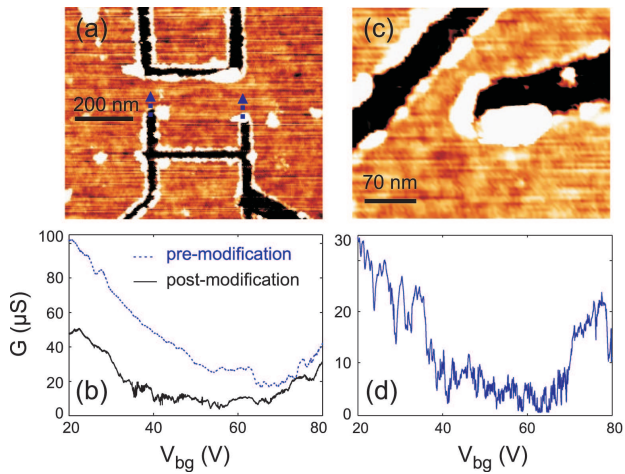}
\caption{\label{Fig3} (color online) (a) AFM image of a quantum dot formed in bilayer graphene using AFM nanolithography. The entrances are
initially $\sim$ 150 nm. In a second AFM lithography step these are then modified, as indicated by the blue
arrows, to be $\sim$ 50 nm. (b) Conductance, G, versus V$_{bg}$ at $T = 4.2$ K of the quantum dot
pre-modification (blue dashed curve) and post-modification (black solid curve). (c) AFM image of a $\sim$ 65 nm constriction
formed in a bilayer flake (d) Conductance versus V$_{bg}$ for the constriction measured
at $T = 4.2$ K.}
\end{figure}

Finally, we show that the technique described is well suited for
device fabrication. Fig.~\ref{Fig3}(a) and \ref{Fig3}(c) show AFM
images of two graphene nanodevices, designed as quantum dot and
quantum wire, respectively, formed in a BLG by AFM nanolithography.
For each device, the tip current is monitored during lithography to
ensure that the graphene is properly cut. Figures.~\ref{Fig3}(b) and
\ref{Fig3}(d) show the conductance as a function of back gate
voltage V$_{bg}$ of the devices at a temperature of 4.2 K. The
flexibility of AFM lithography is illustrated by Fig.~\ref{Fig3}(b)
which shows measurements of the conductance as a function of
V$_{bg}$ at 4.2 K both for the quantum dot as shown in
Fig.~\ref{Fig3}(a) (blue dashed line) as well as that of the same
device but with the entrance barriers of the quantum dot narrowed
from $\sim$ 150 nm to about 50 nm in a subsequent AFM lithography
step at room temperature (black solid line). As expected, the
conductance is significantly lower in the post modification device
with an increase in the gap observed \cite{Han2007}.

In conclusion, we have studied the local oxidation of graphene by an
AFM tip. We demonstrate that at low tip voltages the graphene is
typically not cut even though clear indentations are observed in AFM
height images. The lattice is only cut when the local electric field
exceeds a threshold at which point tip current vanishes (within our
noise floor). These conclusions are supported by scanning electron
microscopy and transport experiments. The ability to distinguish
between pseudo-cuts and cuts as demonstrated here is important for
reliable graphene device fabrication by AFM nanolithography.

We acknowledge Cinzia Casiraghi for technical support. ACF and AL
acknowledge funding from EU grants NANOPOTS and RODIN, EPSRC
EP/G042357/1 and a Royal Society Wolfson Research Merit Award. MRB
acknowledges support from the Royal Society.


\begin{thebibliography}{10}
\bibitem{tips}
The AFM tips are highly doped PointProbe Si tips with a force
constant of 42 N/m. Estimated forces during lithography are in the
1-10 $\mu$N range.
\bibitem{resolution}
The measurement resolution in Fig.~\ref{Fig1} is limited to $\sim
100$ nA. Better resolved measurements (not shown) show that during
AFM lithography I$_{tip} <$ 1 nA.
\end{thebibliography}
\end{document}